\documentclass[twocolumn,prb,aps,showpacs,superscriptaddress]{revtex4-1}
\usepackage{graphicx}
\usepackage[dvipdfm]{hyperref}
\usepackage{amsmath}
\usepackage{epsfig}

\begin{document}
\title{Anisotropy of the molecular magnet V$_{15}$  spin Hamiltonian detected by high-field electron spin resonance}

\author{M. Martens}
\email{martens@magnet.fsu.edu} 
\affiliation{Department of Physics and The National High Magnetic Field Laboratory, Florida State University, Tallahassee, Florida 32310, USA}

\author{J. van Tol} 
\affiliation{The National High Magnetic Field Laboratory, Tallahassee, Florida 32310, USA}

\author{N.S. Dalal}
\affiliation{Department of Chemistry \& Biochemistry, Florida State University, Tallahassee, Florida 32306, USA}

\author{S. Bertaina}
\affiliation{Aix-Marseille Universit\'{e}, CNRS, IM2NP UMR7334, 13397 cedex 20, Marseille, France.}

\author{B. Barbara}
\affiliation{Institut N$\acute{\textrm e}$el, CNRS, Universit$\acute{\textrm e}$ Joseph Fourier, BP 166, F-38042 Grenoble Cedex 9, France}

\author{B. Tsukerblat}
\affiliation{Department of Chemistry, Ben-Gurion University of the Negev, 84105 Beer-Sheva, Israel}

\author{A. M$\ddot{\textrm u}$ller}
\affiliation{Fakult$\ddot{\textrm a}$t f$\ddot{\textrm u}$r Chemie, Universit$\ddot{\textrm a}$t Bielefeld, Postfach 100131, D-33501 Bielefeld, Germany}

\author{S. Garai}
\affiliation{Fakult$\ddot{\textrm a}$t f$\ddot{\textrm u}$r Chemie, Universit$\ddot{\textrm a}$t Bielefeld, Postfach 100131, D-33501 Bielefeld, Germany}

\author{S. Miyashita}
\affiliation{Department of Physics, Graduate School of Science, The University of Tokyo, 7-3-1 Hongo, Bunkyo-ku, Tokyo 113-8656, Japan}

\author{I. Chiorescu}
\email{ic@magnet.fsu.edu} 
\affiliation{Department of Physics and The National High Magnetic Field Laboratory, Florida State University, Tallahassee, Florida 32310, USA}

\date{\today}%

\begin{abstract}
The molecular compound K$_6$[V$^{IV}_{15}$As$^{III}_6$O$_{42}$(H$_2$O)] $\cdot$ 8H$_2$O, in short V$_{15}$, has shown important quantum effects such as coherent spin oscillations. The details of the spin quantum dynamics depend on the exact form of the spin Hamiltonian. In this study, we present a precise analysis of the intramolecular interactions in V$_{15}$. To that purpose, we performed high-field electron spin resonance measurements at 120~GHz and extracted the resonance fields as a function of crystal orientation and temperature.  The data are compared against simulations using exact diagonalization to obtain the parameters of the molecular spin Hamiltonian.
\end{abstract}

\pacs{36.40.Cg,76.30.-v,33.35.+r}

\maketitle

\section{Introduction}

\begin{figure}
\includegraphics[width=3.25 in]{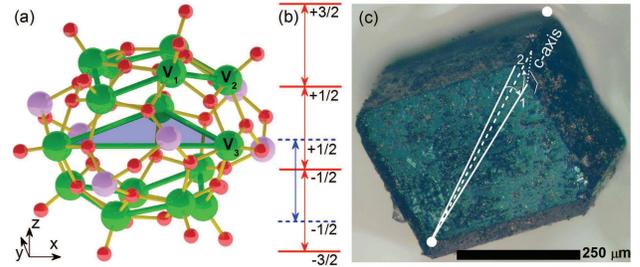}
\caption{(Color online) (a) Ball-and-stick representation of the V$_{15}$ cluster anions, with the $s=1/2$ V ions in green. The $x$ axis is parallel to one of the triangle's sides, and the $z$ axis represents the $c$ axis of the crystal unit cell. (b) Approximate sketch of the energy levels in a field corresponding to a resonant frequency of 120 GHz. Arrows represent the allowed transitions and the numbers are the $S_z$ values. The dashed blue lines indicate the two $S=1/2$ doublets, and the solid red lines show the $S=3/2$ quartet. The resonance fields are averaged in Eq.~(\ref{B1}). (c) V$_{15}$ crystal with markings used to find the $c$ axis in a two-step process (see text). }
\label{fig1}
\end{figure}

The control of the dynamics of spins in solid state materials has direct implications at fundamental and applied levels. Research topics, in particular quantum computing, rely heavily on complex control techniques and long spin coherence times to achieve robust information control. In this context, molecular magnets have gained significant attention in recent years, both for their technological potential and as a test bed for quantum mechanics at a macroscopic scale. The compound K$_6$[V$^{IV}_{15}$As$^{III}_6$O$_{42}$(H$_2$O)] $\cdot$ 8H$_2$O, in short V$_{15}$, has a magnetic cluster anion \cite{Mueller_Angew1988} [Fig.~\ref{fig1}(a)] and it was among the first to show coherence via direct measurements of Rabi oscillations \cite{Sylvain,Shim} as well as interesting out-of-equilibrium spin dynamics due to phonon bottlenecking \cite{IC1,IC2,IC3,Lei1,Lei2}. The observation of spin coherence in molecular magnets is currently not limited to V$_{15}$; it has also recently been studied in systems like Cr$_7$Ni, Fe$_8$, and Fe$_4$, showing the high interest that these systems have gained\cite{Takahashi,Schlegel,Ardavan}. In this work, we study the spin Hamiltonian of V$_{15}$ by means of electron spin resonance (ESR) on a precisely oriented single crystal. We study the Dzyaloshinsky-Moriya (DM) coupling between a molecule's spins as well as the anisotropy of their exchange coupling. Finally, we find that the anisotropy of the $g$-tensor shows an interesting temperature dependence which may indicate the onset of short-range correlations as seen in other systems \cite{Lawrence}. These Hamiltonian parameters are given here with a high degree of precision. The spin Hamiltonian is instrumental to obtain the transition probabilities between certain levels of this multi-level quantum system and thus to predict the spin dynamics at low temperatures.

The V$_{15}$ cluster anions are arranged in a crystal with $R\bar 3c$ symmetry. Each V ion has a spin $1/2$, and the 15 spins are arranged in two hexagons sandwiching a triangle. The couplings between ions V$_1-\textrm V_2$ and V$_3-\textrm V_{1,2}$ are antiferromagnetic and are of the order of hundreds of degrees kelvin \cite{Gatteschi,Barra}. The coupling $J$ between the triangle's corners is also antiferromagnetic \cite{IC2} and can be indirect (via the two hexagons). Therefore, the effective Hamiltonian of the molecule corresponds to that of three spins $s=$1/2 coupled antiferromagnetically \cite{IC2,IC3}. A more precise form of the Hamiltonian is

\begin{multline}
H=\sum_{<i,j>}[\vec D_{ij}(\vec S^{(i)}\times\vec S^{(j)})+ J_zS_z^{(i)}S_z^{(j)}+ J_t(S_x^{(i)}S_x^{(j)}\\
+S_y^{(i)}S_y^{(j)})]+g_a\mu_B\vec B_\triangle \vec S_\triangle + g_c\mu_B\vec B_z \vec S_z
\label{H1}
\end{multline}
where one sums over the $(i,j)$ pairs $(1,2),(2,3),(3,1)$ of triangle spins $\vec S^{(1),(2),(3)}$. The first term represents the DM coupling, and the second one shows the antiferromagnetic coupling between the triangle's corners which, in principle, can be anisotropic, $J_z\ne J_t$. The Zeeman part shows the anisotropy of the $g$ tensor, with $g_{a,c}$ being the values in the triangle's plane and along the molecule's $c$ axis [$z$ axis in Fig.~\ref{fig1}(a)] respectively; $\vec B_{\triangle,z}$ are the planar and vertical vector components of the external magnetic field, respectively, $\vec S_{\triangle}=\vec S_x+\vec S_y= \sum_{i=1,2,3}(\vec S^{(i)}_x+\vec S^{(i)}_y) $, $\vec S_{z}=\sum_{i=1,2,3}\vec S^{(i)}_z$, and $\mu_B$ is the Bohr magneton. The resulting eigenvalues describe an excited quartet $S=3/2$ and two ground doublets $S=1/2$. The DM coupling, although usually considered to be very small, can be sufficient to lift the degeneracy of the two doublets and, in second order, the zero-field degeneracy of the $S=3/2$ quartet (see theoretical Refs. \onlinecite{Tsukerblat_JCP2006,Boris2,Tarantul,Machida,Seiji,Uchiyama} for more details).

\section{Experimental methods}

The continuous-wave ESR experiments were performed at 120 GHz on the quasioptical superheterodyne spectrometer at the National High Magnetic Field Laboratory \cite{Morley,van Tol}. The magnet used is a sweepable 12.5-T superconducting magnet with a homogeneity of $10^{-5}$ over 1 cm$^3$. The experimental setup allows for a continuous change of the angle $\theta=\arctan(B_\triangle/B_z)$ between the $c$ axis of the molecule and the static field $\vec B_0=\vec B_\triangle +\vec B_z$. The temperature can be varied from room temperature down to 2.5~K. The field is scanned and when the induced Zeeman splittings are equal to 120~GHz, resonant absorption of the microwaves is observed in the power reflected by the crystal. The measured resonant fields are compared against values computed using the method of first moments \cite{Tsukerblat_JCP2006}. There are three transitions within the quartet [shown by red arrows in Fig.~\ref{fig1}(b)] and four within the two doublets whose fields are found by exact diagonalization of the three-spin Hamiltonian (\ref{H1}). Using transition probabilities as weights, these fields are thermally averaged to produce a temperature-dependent resonance field:
\begin{equation}
B=\sum_{i=1}^{7}I_iB_i/\sum_{i=1}^{7}I_i
\label{B1}
\end{equation}
with
\begin{equation}
I_i=\frac{|\langle b|\vec S_\phi|a\rangle|^2}{Z(B_i)}\left[\exp\left(-\frac{E_a}{k_BT}\right)- \exp\left(-\frac{E_b}{k_BT}\right)\right].
\end{equation}
Here $Z(B_i)$ is the partition function over the eight levels at fixed field $B_i$ and $E_{a,b}$ are eigenvalues for the states $|a\rangle$ and $|b\rangle$ involved in a given transition, that is, $E_a-E_b=hF_{mw}$, with $F_{mw}=$120~GHz and $k_B$ and $h$ being the Boltzmann and Planck constants, respectively. The operator $\vec S_\phi=\vec S_x\sin\phi-\vec S_y\cos\phi$ represents the microwave drive $\vec h_{mw}\perp \vec B_0$ (here $\phi$ is the angle between $\vec B_\triangle$ and the $x$ axis).

\begin{figure}
\includegraphics[width=3.37 in]{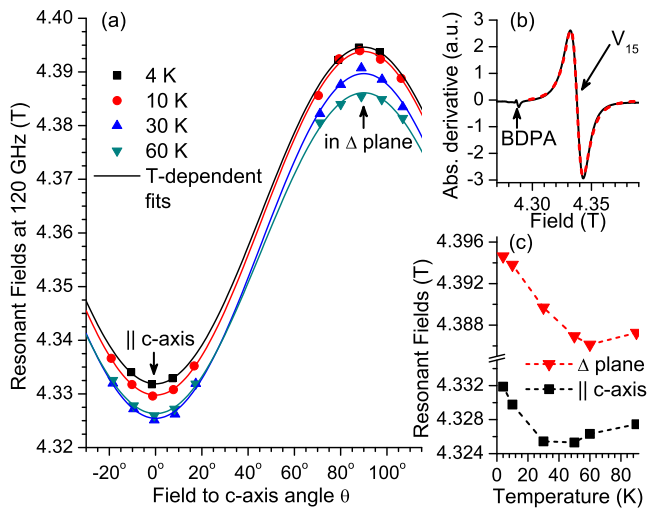}
\caption{(Color online) (a) Angular dependence of the resonance field as a function of the angle $\theta$ with the $c$ axis, at four different temperatures. The solid lines show fits using Eq.~(\ref{B1}). (b) Example of ESR spectrum showing the first derivative of the absorption signal (in arbitrary units, a.u.) of the calibration standard (BDPA) and the molecule at 10~K. The dashed line shows the fit used to obtain the resonance field. (c) Variation of the minimum $B_{min}$ ($\theta=0^{\circ},B_0||c$) and maximum $B_{max}$ ($\theta=90^{\circ},B_0$ in the triangle plane) of the resonance fields with temperature.}
\label{fig2}
\end{figure}

The experiments are performed on a single crystal, shown in Fig.~\ref{fig1}(c), with typical, well-defined facets. Such facets are not unusual among $R\bar 3c$ compounds; for example, calcite has a similar exterior aspect. It is of essence to be able to correctly identify the $c$ axis of the unit cell (in hexagonal representation \cite{Barra}) since the V single-site anisotropy is translated at the level of the entire molecule \cite{Gatteschi}. Thus, the $g$ factor measured along the $c$ axis has the maximum value ($g_c=g_\perp$), while in the triangle plane, it is averaged between the single-site directional values $g_{\perp,||}$ [$g_a=\sqrt{(g_\perp^2+g_{||}^2)/2}$, see Ref.~\onlinecite{Gatteschi} for details]. This information allows us to find the $c$ axis through a process of rotating the crystal in the applied magnetic field and recording the ESR spectrum (reflected power vs field) at each rotational step $\theta$. After a full rotation, the crystal is extracted and repositioned on the rotator stage. For each new position, the resonance field vs $\theta$ shows an oscillatory behavior [like in Fig.~\ref{fig2}(a)] with a certain amplitude. The repositioning is iterated until a maximum amplitude is obtained. The $c$ axis is parallel to the applied field when the minimum resonance field occurs, and $\theta$ is thus calibrated as $\theta=0^\circ$ with an uncertainty of $\sim 1^\circ$. Figure~\ref{fig1}(c) shows the final location of the $c$ axis. Since the crystal shape is common among V$_{15}$ crystals, the result can be generalized to quickly find the $c$ axis using a two-step process followed by fine-tuning using the iterative procedure above. In step 1, one looks straight down at the crystal, and a line is drawn at a 2.6$^\circ$ angle to the short diagonal clockwise and along the top face plane. The pivot point of this rotation must be the corner that overhangs the bottom of the crystal [shown with white dots in Fig.~\ref{fig1}(c)]. In step 2, the line is rotated 42$^\circ$ directly into the crystal; this rotation is exactly orthogonal to the first one. The final direction, shown by the dashed line starting at the overhanging corner of the short diagonal, is parallel to the $c$ axis of the unit cell.

Using the procedure described above, one can perform $\theta$ rotations in a plane containing the $c$ axis and identify the resonance field for each angle. Also, the magnetic field is found by using the well-known ESR calibration standard 1,3-bisdiphenylene-2 phenylallyl (BDPA \cite{bdpa}, $g=2.00263$) placed on the same sample holder as the V$_{15}$ crystal. The measurements are repeated for temperatures ranging from 3 to 90 K. The data shown in Fig.~\ref{fig2}(a) correspond to three such temperatures (dots), while the continuous fits are numerical simulations allowing one to identify the Hamiltonian parameters (see below). The resonance field shows a minimum $B_{min}$ when the field is $||c$ and reaches a maximum $B_{max}$ when the field is in the triangle ($\triangle$) plane. A typical ESR spectrum is shown in Fig.~\ref{fig2}(b) as the first derivative of the absorption signal vs field at 10~K. At a lower field one observes the signal from the BDPA standard, while the V$_{15}$ signal is larger and placed at higher fields. The dashed red line shows the fit used to extract the resonance fields. The fit uses a typical ESR model which sums up the absorption and dispersion components of the signal\cite{Poole} since there is no automatic frequency control to cancel the dispersion part. Note that the width of a typical ESR line is of the order of few tens of milliteslas (most likely due to hyperfine and dipolar interactions), thus allowing a precise fitting for resonance field identification with an uncertainty of $\sim$0.5~mT. The temperature dependence of $B_{min}$ and $B_{max}$ is shown in Fig.~\ref{fig2}(c) and is an essential result used below to discuss the $J$ and DM couplings.

\begin{figure}
\includegraphics[width=3.37 in]{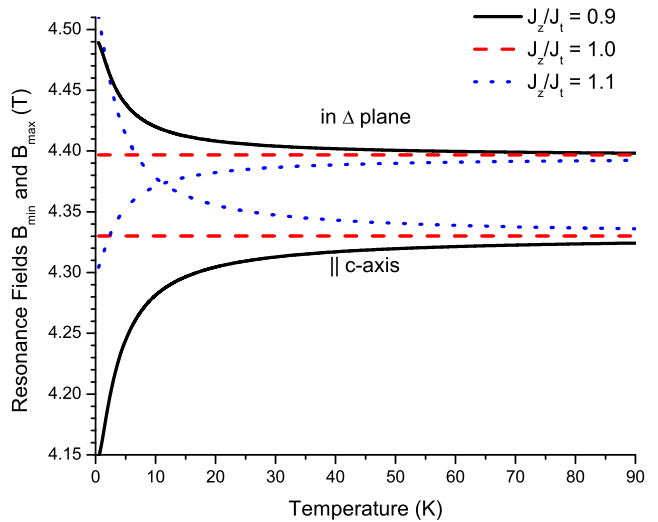}
\caption{(Color online) The resonance fields along the c-axis ($B_{min}$) and $\perp$ to it ($B_{max}$) as a function of temperature for an isotropic J coupling (dashed line) and two anisotropic situations, $J_z$ larger (dotted line) and smaller (line) than $J_t$.}
\label{fig3}
\end{figure} 

\section{Exchange coupling anisotropy}

In Hamiltonian~(\ref{H1}), the antiferromagnetic coupling between spins must lead to a spin crossover \cite{IC2} at 2.8~T when the ground state switches between $S=1/2, S_z=-1/2$ and $S=3/2, S_z=-3/2$. To account for a potential anisotropy of this coupling, we consider $J_t=-2.45$~K and the ratio $J_z/J_t$ to be variable (the conclusions of this discussion are the same if we fix $J_z$ instead of $J_t$). We can observe that such anisotropy can, in principle, generate a zero-field splitting for the quartet state $S=3/2$. Indeed, by writing this part of the Hamiltonian as $J_t\vec S^{(i)}\vec S^{(j)}+ (J_z-J_t)S_z^{(i)}S_z^{(j)}$ one can observe that $J_z-J_t$ plays a role similar to that of a crystal field term $DS_z^2$, thus potentially raising the degeneracy between the $S_z=\pm 3/2$ and $\pm 1/2$ states of the quartet. Using Eq.~(\ref{B1}), we simulate the temperature dependence of $B_{min,max}$ for some typical values of the $g$ factors \cite{Sylvain}, namely, $g_a=$1.95 and $g_c=$1.98. The qualitative behavior shown in Fig.~\ref{fig3} is compared to that in Fig.~\ref{fig2}(c) and we find that for $J_z < J_t$ ($J_z>J_t$), $B_{min}$ ($B_{max}$) contradicts the experimental behavior by increasing instead of decreasing with increasing temperature. If the ratio $J_{z}/J_{t}$ is sufficiently large, a crossing between $B_{max}$ and $B_{min}$ is observed. A crossing is measured in V$_{15}$ at very high temperatures (200~K) \cite{Ajiro} when all 15 spins of the molecule contribute to its magnetism. In contrast, our study deals with the low-temperature regime where the three-spin model is adequate. Such a crossing would imply a large increase in amplitude $B_{max}-B_{min}$ at very low temperatures, whereas, experimentally, the amplitude is relatively constant. Consequently, to properly explain the data we consider $J_{z,t}=-2.45$~K (dashed line).

\begin{figure}
\includegraphics[width=3.37 in]{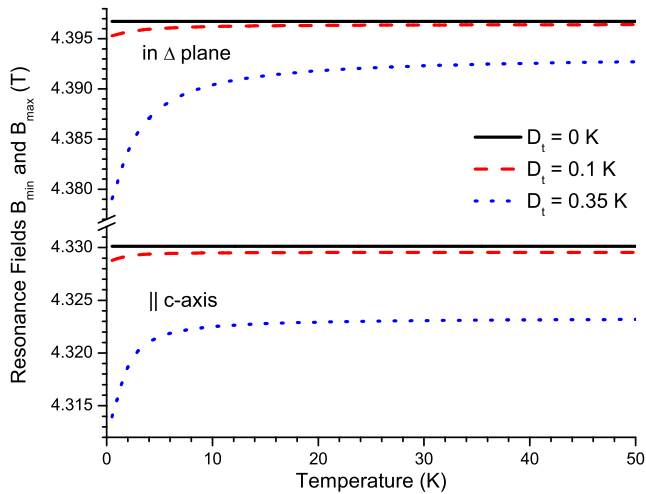}
\caption{(Color online) The resonance fields along the $c$ axis ($B_{min}$) and perpendicular to it ($B_{max}$) as a function of temperature for different values of the DM vector.}
\label{fig4}
\end{figure} 

\section{Dzyaloshinsky-Moriya interaction}

The DM term of Hamiltonian~(\ref{H1}) can mix states with different total spins and can lift the degeneracy of the two doublets. The DM vector components are considered to be the same for all three spin pairs and are defined as in Ref. \onlinecite{Tsukerblat_JCP2006}, with $D_z$ being the vertical component and $D_t$ being the effective in-plane DM interaction incorporating both components, perpendicular and along the side of the triangle. For instance, the spin crossing at 2.8~T can show a level repulsion if a $D_t$ term is present, while a $D_z$ term is effective in raising the doublets degeneracy. Consequently, there is a potential to split and identify all the resonance fields and their transition probabilities that enter Eq.~(\ref{B1}).  

We studied the effect of the DM coupling on the value of the resonance fields and the results are presented in Fig.~\ref{fig4} for $D_t=0, 0.1, 0.35$~K. The second value (0.1~K) is somewhat large when compared to previous studies \cite{IC3} based on low-temperature magnetization reversal curves (DM estimated at 0.05~K) in order to show its effect. The last one (0.35~K) is very large for this molecule to exaggerate the effect on the resonance fields. The simulations show a small effect but opposite to the measured behavior [see Fig.~\ref{fig2}(c)] in that the resonance fields are increasing with temperature instead of decreasing. The effect is orders of magnitude smaller if one replaces $D_t$ with $D_z$ (not shown). Therefore, we conclude that $D_t$ has to be sufficiently small (up to $\sim$10 mK) or even nonexistent such that its effect on resonance fields is negligible (since it is opposing the observed behavior). A similar conclusion is drawn for the $D_z$ component or for a situation when the exchange couplings between the triangle's corners are not identical \cite{Chaboussant_epl_2004}. Both of these cases would generate satellite transitions, and we do not observe such signals within the experiment resolution (a few milliteslas). This behavior and the discussed effect of an anisotropic coupling $J_z\neq J_t$ persist for any angle $\phi$. Consequently, we consider in the following no DM coupling and isotropic $J$, in which case the seven resonance fields are all equal and varying the angle $\phi$ has no effect on the resonance fields. Thus, we can further simplify the Hamiltonian by taking $\phi=0$. 

\begin{figure}
\includegraphics[width=3.37 in]{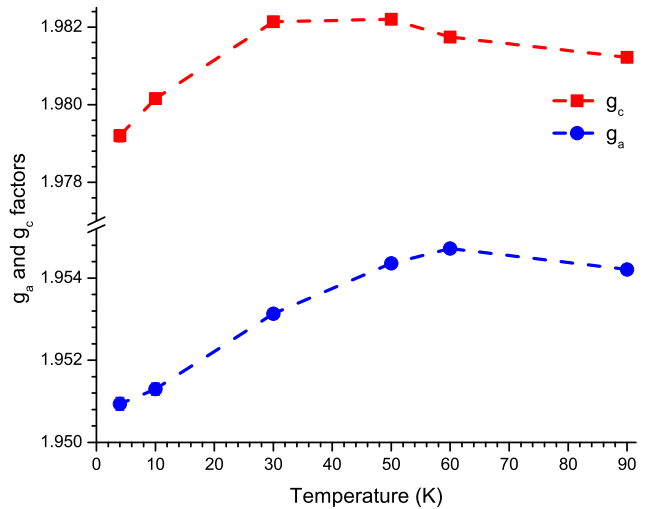}
\caption{(Color online) Values of the $g_a$ and $g_c$ $g$ factors obtained from the angular dependences shown in Fig.~\ref{fig1} for seven temperatures. The error bars are smaller than the symbol size. The $g$ factors increase with the temperature by $\sim 0.2\%$.  }
\label{fig5}
\end{figure} 
\begin{table}
\begin{center}
\begin{tabular}{|c|c|c|c|c|}\hline
Temperature (K) & $g_a$ & $\sigma_{g_a}(\times 10^4)$ & $g_c$ & $\sigma_{g_c}(\times 10^4)$  \\ \hline
4 & 1.9509 & 1.6 & 1.9792 & 1.6  \\ \hline
10 & 1.9513 & 1.5 & 1.9802 & 1.1  \\ \hline
30 & 1.9531 & 1.2 & 1.9821 & 1  \\ \hline
50 & 1.9544 & 1.2 & 1.9822 & 1.2  \\ \hline
60 & 1.9547 & 1 & 1.9817 & 1.2  \\ \hline
90 & 1.9542 & 1 & 1.9812 & 1  \\ \hline
\end{tabular}
\caption{Values of the $g_a$ and $g_c$ $g$ factors shown in Fig.~\ref{fig5}.}
\label{Table1}
\end{center}
\end{table}

\section{Anisotropy of the $\textrm{g}$-tensor}

Using a value of $J_z=J_t=2.45$~K, universal constant values as recommended by the 2010 CODATA database, and the field calibrated with BDPA as explained above, we performed the fits presented in Fig.~\ref{fig2} with solid lines. To determine the anisotropy of the $g$ tensor at different temperatures, the $\chi^2$ parameter is minimized. The fitted $g$ factor values have a precision of four decimal places. 

We note that the $g$ factor increases only slightly, by 0.2$\%$ when the temperature is increased (see Table~\ref{Table1} and Fig.~\ref{fig5}). Thermal variations in the $g$ factor are sometimes quite sharp, indicating magnetic or structural transitions, as is the case for some organic conductors \cite{Tokumoto_PRL2008} or for Cu and Mn spins in NH$_4$Cl (see Refs. \onlinecite{Pilbrow_PhysStatSol_1967,Watanabe_PhysLett_1975,Stibbe_Physica_1978}). The decrease of temperature can alter the local symmetry of $3d$ orbitals in the V ions, thus altering their energy splittings. Using first- and second-order perturbation theory, this is demonstrated \cite{Pryce_ProcRSoc_1950} to induce a variation of the anisotropic $g$ tensor in the case of anisotropic orbitals.

The smooth decrease in the $g$ components below $\sim$30-40~K is suggestive of changes in the  molecular structure of the V$_{15}$ anion at a local length scale or of an ensuing phase transition involving a long-range lattice order. No earlier study down to 30 mK has observed any long-range ordering  (see, for example, Ref.~\onlinecite{IC3}). The observed $g$ value thus clearly indicates that there is a change in the local molecular geometry which could be due to the onset of short-range correlations in the lattice. Our report on the possibility of such local ordering might be of significance in understanding the spin dynamics below 30 K.  

Also, it is important to note that the effect of a small DM coupling can easily be absorbed into a small decrease of the $g$ factor. For instance, even if one considers a very large DM vector of size 0.35~K, the $g$ factors would have to decrease by only $\le 0.4\%$ to accommodate it (not shown). As mentioned above, such large values of the DM vector would not be justified, and any DM coupling would induce effects opposed to the measured temperature dependence of the resonance fields. Consequently, it is more suitable to consider the DM vector as being equal to zero and to use the $g$ factor values given in Table \ref{Table1}.

\section{Conclusions}

In conclusion, we present a determination of the spin Hamiltonian parameters of V$_{15}$, a prototypical molecular magnet which has gained a large amount of interest recently. Angular-variation and temperature-dependence studies have been carried out on a precisely oriented single crystal. These measurements show that the anisotropic component of the DM vector is $<10^{-2}$~K, smaller than earlier estimates. Also observed are significant changes in the $g$ tensor at temperatures below 30 K, possibly indicating local order. This information should be of great interest for understanding the low-temperature spin dynamics and electronic structure and bonding in V$_{15}$. 

\acknowledgements

We thank Dr. H. B\"ogge (Bielefeld) for some crystallographic remarks. This work was supported by NSF Grant No. DMR-1206267 and CNRS-PICS CoDyLow. The NHMFL is supported by Cooperative Agreement Grant No. DMR-0654118 and the state of Florida. Partial support was received from JSPS KAKENHI Grant-in-Aid for Scientific Research (C) No. 25400391 from MEXT of Japan, and some numerical calculations were supported by the supercomputer center of ISSP of the University of Tokyo. B.T. acknowledges financial support of the Israel Science Foundation (ISF Grant no. 168/09) and COST Action CM1203 \textquotedblleft Polyoxometalate Chemistry for molecular Nanoscience\textquotedblright  (PoCheMon).


\begin{thebibliography}{99}
\bibitem{bdpa}The $g$ value of unoriented BDPA in polystyrene (0.08 wt \%) was calibrated by high-field electron nuclear double resonance at 240 and 411 GHz and was found to be $g=(2g_\perp+g_{||})/3=2.00263$, with $g_\perp=2.00271(5)$ and $g_{||} = 2.00246(5)$. These values are considered to be temperature independent.
\end{thebibliography}
\end{document}